\documentclass[twocolumn,showpacs,superscriptaddress,aps,prl]{revtex4-1}
\usepackage{graphicx}
\usepackage[caption=false,subrefformat=parens,labelformat=parens]{subfig}
\usepackage{hyperref}
\hypersetup{
    colorlinks=true,       
    linkcolor=blue,        
    citecolor=blue,        
    filecolor=blue,        
    urlcolor=blue          
}

\begin{document}

\title{Ferro-orbital ordering transition in iron telluride Fe$_{1+y}$Te}

\author{David Fobes}
\email{dfobes@bnl.gov}
\affiliation{CMPMSD, 
 Brookhaven National Laboratory, Upton, NY 11973 USA}

\author{Igor A.\ Zaliznyak}
\email{zaliznyak@bnl.gov}
\affiliation{CMPMSD, 
 Brookhaven National Laboratory, Upton, NY 11973 USA}

\author{Zhijun Xu}
\affiliation{CMPMSD, 
 Brookhaven National Laboratory, Upton, NY 11973 USA}

\author{Ruidan Zhong}
\affiliation{CMPMSD, 
 Brookhaven National Laboratory, Upton, NY 11973 USA}

\author{Genda Gu}
\affiliation{CMPMSD, 
 Brookhaven National Laboratory, Upton, NY 11973 USA}

\author{John M. Tranquada}
\affiliation{CMPMSD, 
 Brookhaven National Laboratory, Upton, NY 11973 USA}

\author{Leland Harriger}
\affiliation{NCNR, 
 National Institute of Standards and Technology, Gaithersburg, MD 20899 USA}

\author{Deepak Singh}
\affiliation{NCNR, 
 National Institute of Standards and Technology, Gaithersburg, MD 20899 USA}

\author{V. Ovidiu Garlea}
\affiliation{QCMD, 
 Oak Ridge National Laboratory, Oak Ridge, TN 37831 USA}

\author{Mark Lumsden}
\affiliation{QCMD, 
 Oak Ridge National Laboratory, Oak Ridge, TN 37831 USA}

\author{Barry Winn}
\affiliation{QCMD, 
 Oak Ridge National Laboratory, Oak Ridge, TN 37831 USA}

\begin{abstract}
Fe$_{1+y}$Te with $y \lesssim 0.05$ exhibits a first-order phase transition on cooling to a state with a lowered structural symmetry, bicollinear antiferromagnetic order, and metallic conductivity, $d\rho/dT > 0$. Here, we study samples with $y = 0.09(1)$, where the frustration effects of the interstitial Fe decouple different orders, leading to a sequence of transitions. While the lattice distortion is closely followed by \emph{incommensurate} magnetic order, the development of \emph{bicollinear} order and metallic electronic coherence is uniquely associated with a separate hysteretic first-order transition, at a markedly lower temperature, to a phase with dramatically enhanced bond-order wave (BOW) order. The BOW state suggests ferro-orbital ordering, where electronic delocalization in ferromagnetic zigzag chains decreases local spin and results in metallic transport.
\end{abstract}


\maketitle

In a pattern common with cuprates, iron pnictide and chalcogenide superconductors (FeSC) have parent phases, which, upon cooling, undergo antiferromagnetic (AFM) ordering and structural distortion(s) lowering the high-temperature tetragonal (HTT) paramagnetic lattice symmetry \cite{Scalapino2012,DaiHuDagotto2012}. They also host strong magnetic fluctuations, a hallmark of unconventional superconductivity \cite{LumsdenChristianson2010}. Recently, there has also been strong experimental evidence of broken electronic symmetry, ``nematicity'', accompanying, or preceding the magnetic/lattice ordered phase, reminiscent of stripes in cuprates \cite{Fisher_RPP2011}. The physics driving these phenomena, their inter-relation and relation to the superconductivity remain unclear \cite{DaiHuDagotto2012}.

\begin{figure}[t!pb]
\subfloat{
\includegraphics[width=0.99\linewidth]{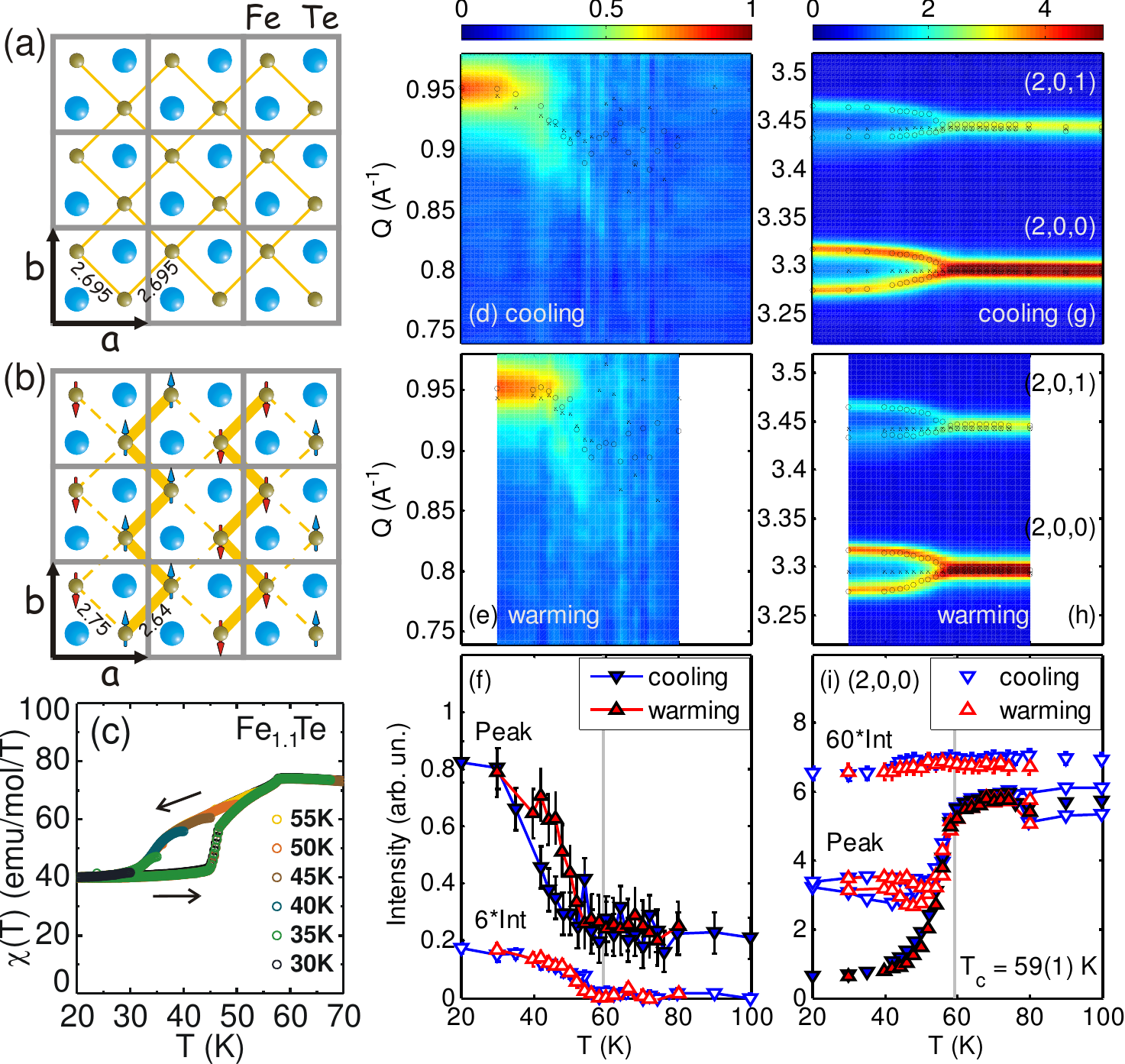}
\label{fig0:structure:a}}
\subfloat{\label{fig0:structure:b}}\subfloat{\label{fig0:structure:c}}\subfloat{\label{fig0:structure:d}}\subfloat{\label{fig0:structure:e}}\subfloat{\label{fig0:structure:f}}\subfloat{\label{fig0:structure:g}}\subfloat{\label{fig0:structure:h}}\subfloat{\label{fig0:structure:i}}\subfloat{\label{fig0:structure:j}}\subfloat{\label{fig0:structure:k}}
\vspace{-0.1in}
\caption{$P4/nmm$ unit cell of the square-lattice structure of FeTe in the $a$-$b$ plane, (a), and formation of zigzag chains by atom displacements from high symmetry positions allowed in $P2_{1}/m$ structure, overlain with the bicollinear magnetic structure, (b). Magnetic susceptibility measured on cooling and warming upon zero-field cooling to different initial temperature, (c). Magnetic, (d)--(f), and structural, (g)--(i), intensity measured by neutron powder diffraction on cooling and warming. Peak magnetic intensity in (f) shows hysteresis, while integral magnetic intensity does not. Closed symbols in (i) show the intensity at the (2,0,0) center of mass (COM) position, open symbols with lines show peak intensities on two sides of the COM. Both decrease below $T_{c} \approx 60$~K, indicating peak splitting. Open symbols show the integral intensity of nuclear (2,0,0) peak.}
\label{fig0:structure}
\vspace{-0.2in}
\end{figure}

Unlike cuprates, the Fe-based materials have several unfilled $3d$ bands. Their parent magnetic phases have well-defined Fermi surfaces, indicating a metallic nature \cite{Fisher_RPP2011,Singh2012}. Such ``weak Mott-ness'' and itinerancy, combined with orbital degeneracy entangled with the magnetic and lattice degrees of freedom, leads to the proliferation of theoretical models and approaches: strong coupling where physics is spin-driven \cite{SiAbrahams2009}, weak coupling where it is determined by properties of the electronic Fermi surface \cite{FernandezChubukov_PRB2012}, or mixed spin-orbital models \cite{Kruger2009,LvWuPhillips2009,StanevLittlewood2013}. The experimental evidence enabling one to distinguish among these models is, however, still scarce. Here we present such evidence for the case of Fe$_{1+y}$Te, the end member of the chalcogenide family of FeSC, where correlation effects are the strongest \cite{YinHauleKotliar2011}. By combining the results of bulk characterization of electronic behavior and neutron diffraction data on the temperature evolution of the microscopic structure we are able to disentangle different low-temperature orders and show that the transition to the magnetically-ordered state \cite{Bao2009,Li2009}, illustrated in Fig.~\subref*{fig0:structure:b}, is electronically driven through ferro-orbital ordering of zigzag Fe--Fe chains.

The iron-chalcogenides Fe$_{1+y}$Te$_{1+x}$Se$_{x}$, with $T_{c}\approx14.5$~K at optimal doping, consist of a continuous stacking of Fe square-lattice layers, separated by two half-filled chalcogen (Te,Se) layers \cite{Yeh2008,Hsu2009,Wen2011}. Predicted by band structure calculations to be a metal \cite{Singh2012}, non-superconducting parent material Fe$_{1+y}$Te shows non-metallic character in resistivity at high temperature, $d\rho/dT < 0$, indicative of charge carrier incoherence near the Fermi level \cite{LiuPRB2009,Chen2009,HuPetrovic2009}. Curie-Weiss behavior of magnetic susceptibility reveals large, $\sim$4$\mu_{B}$ local magnetic moments, indicating full involvement of three electronic bands \cite{Chen2009,HuPetrovic2009,Zaliznyak_PRB2012}. Nevertheless, angle-resolved photoemission (ARPES) studies show significant spectral weight near the Fermi energy \cite{Xia2009,ZhangFeng2010,Liu_Shen2013}.

The electronic and magnetic properties of Fe$_{1+y}$Te are extremely sensitive to non-stoichiometric Fe at interstitial sites \cite{McQueen_PRB2009,Bao2009,Rodriguez2011,Li2009,Martinelli2010,Stock2011,Rossler2011,Liu2011}. At low concentrations, $y \lesssim 0.05$, there is a first-order magnetostructural transition from the paramagnetic tetragonal $P4/nmm$ phase to monoclinic $P2_{1}/m$ with bicollinear AFM order with propagation vector $q=(0.5,0,0.5)$ and metallic resistivity \cite{Bao2009,Li2009,Martinelli2010,Rodriguez2011}. At high $y \gtrsim 0.12$, the low-T phase is orthorhombic $Pnmm$, with a strongly incommensurate helimagnetic spin order and the resistivity remains semiconducting. The interstitial Fe frustrates bicollinear antiferromagnetism and introduces random lattice strain, which hinder lowering of the HTT symmetry \cite{Zaliznyak_PRB2012,Zaliznyak_PRL2011,Thampy2012,ZacharZaliznyak2003}.

Here we study Fe$_{1+y}$Te in the intermediate range, $y = 0.09(1)$, where the low-$T$ phase is bicollinear AFM, common to $y \lesssim 0.12$, but the magneto-structural transition is split into a sequence of transitions. A lattice distortion occurs at $T_{S} = 61(2)$~K and is followed by a slightly incommensurate magnetic order at $T_{N} = 57.5(5)$~K \cite{Zaliznyak_PRB2012}. They are then followed by a transition at a markedly lower temperature and with a significant cooling-warming hysteresis, leading to a substantial decrease in magnetic susceptibility [Fig.~\subref*{fig0:structure:c}], but nearly absent in the heat capacity, indicating only very small magnetic entropy is involved \cite{HuPetrovic2009,Zaliznyak_PRB2012}. This hysteresis has previously been reported as the difference between field-cooled (FC) and zero-field-cooled (ZFC) susceptibilities, suggesting a putative glassy behavior. Here we uncover the true nature of this transition and show it has profound consequences for the magnetic and electronic properties.

Neutron measurements were carried out using the Hybrid Spectrometer (HYSPEC) and the POWGEN diffractometer at the Spallation Neutron Source and the Powder Diffractometer (HB-2A) at the High Flux Isotope Reactor ($E_{i} = 35$~meV) at Oak Ridge National Laboratory, and the Spin Polarized Inelastic Neutron Spectrometer (SPINS) at the NIST Center for Neutron Research ($E_{i} = 5$~meV). At SPINS, a cooled Be filter after the sample was used to minimize intensity at harmonics of the desired wavelength. A single crystal ($m = 18.45$ g) with a mosaic of $2.2^{\circ}$ full width at half maximum (FWHM), grown by horizontal Bridgman method \cite{Wen2011}, was mounted on an aluminum holder. The crystal was aligned with \textit{c}-axis vertical, measuring elastic scattering in the $(H,K,0)$ plane. The powder sample was obtained by grinding a similar single crystalline piece. Magnetic susceptibility and resistivity on a comparable single crystal were measured at Brookhaven National Laboratory using a Quantum Design Magnetic Property Measurement System (MPMS). Sample compositions were determined to be $y=0.09(1)$ using the methods described in Ref. \cite{Zaliznyak_PRB2012}.

Fig.~\ref{fig0:structure} presents a partial overview of the structural and magnetic temperature dependence observed in this system. Fig.~\subref*{fig0:structure:c} shows magnetic susceptibility data on a small single crystal sample with different ZFC/FC histories, performed by varying the temperature at which FC was initiated, illustrating that the observed hysteresis in the 30--50~K range does not result from glassy magnetic state because it is field-history-independent. In Figs.~\subref*{fig0:structure:d}--\subref{fig0:structure:i} we present results of neutron powder diffraction (NPD) data on a sample having similar transition temperatures to our single crystal samples. Figs.~\subref*{fig0:structure:d} and \subref*{fig0:structure:e} focus on the magnetic $q \approx (0.5,0,0.5)$ peak upon cooling and warming, respectively. Note that near the onset of magnetic order, the wave vector is incommensurate, becoming commensurate at lower temperatures. The integral intensity displays no hysteresis, unlike the peak intensity [Fig.~\subref*{fig0:structure:f}]. Figs.~\subref*{fig0:structure:g} and \subref*{fig0:structure:h} show the temperature dependence of the (2,0,0) and (2,0,1) structural peaks. The structural transition is revealed by peak splitting and an abrupt change in peak intensity at $T_{S} \approx 59(1)$~K. Neither the integrated nor peak structural intensities display hysteresis [Fig.~\subref*{fig0:structure:i}].

\begin{figure}[tpb]
\subfloat{
\includegraphics[width=0.95\linewidth]{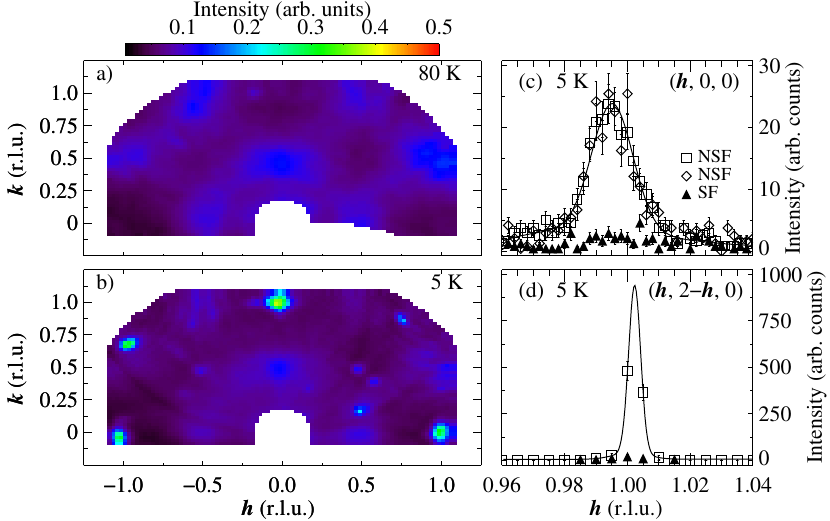}
\label{fig1:hyspec:a}}
\subfloat{\label{fig1:hyspec:b}}\subfloat{\label{fig1:hyspec:c}}\subfloat{\label{fig1:hyspec:d}}
\vspace{-0.1in}
\caption{Elastic neutron scattering from Fe$_{1.09(1)}$Te at 80~K (a), and 5~K (b), measured on HYSPEC ($E_{i} = 7.75$~meV). (1,0,0) and (0,1,0) Bragg peaks, which measure Fe displacements from high-symmetry positions in the $a$-$b$ plane, are seen at 5~K but not at 80~K. Panels (c) and (d) show scans through (1,0,0) and (1,1,0) peaks with polarized neutrons measured at SPINS, confirming the non-magnetic nature of (1,0,0).}
\vspace{-0.2in}
\label{fig1:hyspec}
\end{figure}

In Fig.~\ref{fig1:hyspec} we present elastic neutron scattering maps of our single crystalline sample measured on HYSPEC at 80~K [Fig.~\subref*{fig1:hyspec:a}] and 5~K [Fig.~\subref*{fig1:hyspec:b}]. At 80~K the signal is dominated by broad diffuse magnetic scattering centered around ($\pm 0.5, 0$) and ($0, \pm 0.5$), but at 5~K we observe additional Bragg peaks at (1,0,0) and (0,1,0) positions (equivalent due to the presence of twinning in the crystal), not observed above the structure transition $T_{S} \approx 60$~K. These Bragg reflections are expected in neither the low temperature $P2_{1}/m$ nor high temperature $P4/nmm$ symmetries for atoms in their high symmetry positions, \textit{i.e.} Fe (0.75, 0.25) and Te (0.25, 0.25) \cite{Li2009,supplement}. Additional peaks at 5~K have been identified as spurious multiple scattering by varying the incident energy \cite{supplement}. Polarized neutron scattering  measurements at SPINS for the (1,0,0) peak and the high-symmetry-allowed (1,1,0) [Fig.~\subref*{fig1:hyspec:c} and \subref*{fig1:hyspec:d}] reveal no spin-flip scattering, suggesting that the (1,0,0)/(0,1,0) peak is not a result of magnetic order.
While this reflection is too weak to be observable in NPD, it is consistent with Fe/Te displacements from high-symmetry positions in $P2_{1}/m$ phase, which have been reported in most \cite{Zaliznyak_PRB2012,Bao2009,Martinelli2010,Rodriguez2011}, but not all \cite{Li2009}, previous NPD structural refinements.

\begin{figure}[tpb]
\subfloat{
\includegraphics[width=0.95\linewidth]{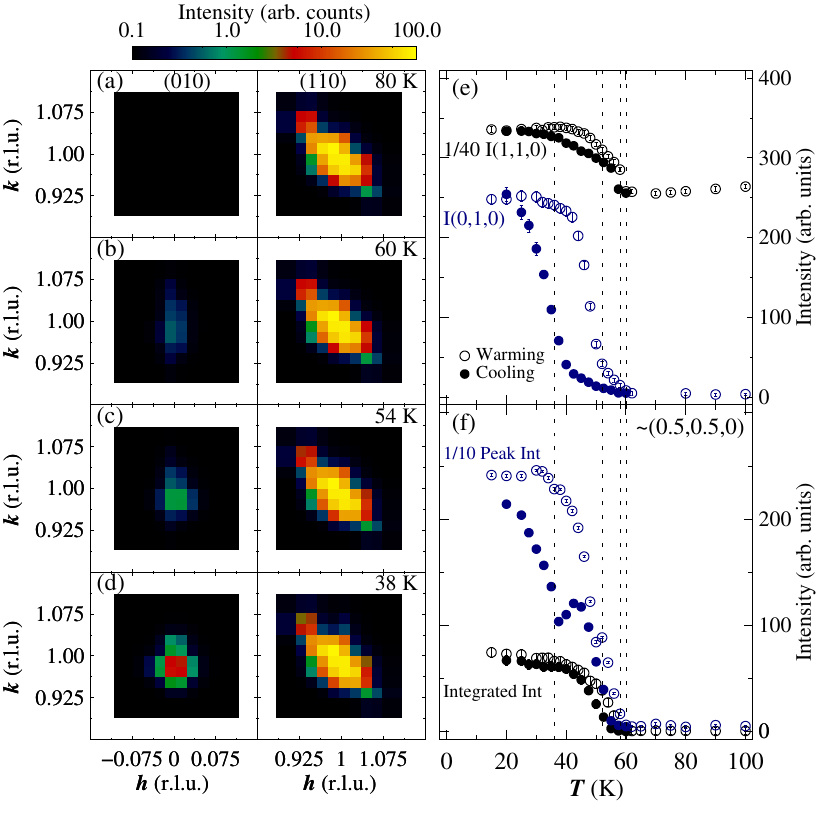}
\label{fig2:tdep:a}}
\subfloat{\label{fig2:tdep:b}}\subfloat{\label{fig2:tdep:c}}\subfloat{\label{fig2:tdep:d}}\subfloat{\label{fig2:tdep:e}}\subfloat{\label{fig2:tdep:f}}
\vspace{-0.1in}
\caption{Intensity map of elastic scattering near (0,1,0) (left) and (1,1,0) (right) measured in Fe$_{1.09(1)}$Te on SPINS ($E_{f} = 5$~meV) at several temperatures, (a)--(d). Intensity of (0,1,0) and (1,1,0) peaks obtained by numerically computing the corresponding moments of the net neutron intensity, on cooling (closed) and warming (open), (e). Peak (hysteretic) and integral (non-hysteretic) intensity of $(0.5-\delta,0,0.5)$ magnetic Bragg intensity, which is seen near (0.5,0.5,0) due to the double scattering in this configuration, (f). Dashed lines, as in other figures, show different transition temperatures.}
\label{fig2:tdep}
\vspace{-0.2in}
\end{figure}

To further investigate the evolution and origin of the (0,1,0) peak, we performed a detailed study of the temperature dependence, presented in Fig.~\ref{fig2:tdep}. Elastic neutron scattering maps of the (0,1,0) and (1,1,0) Bragg peaks at various temperatures [Figs.~\subref*{fig2:tdep:a}--\subref{fig2:tdep:d}] reveal that (0,1,0) is absent at 80~K and present below $T_{S}$. Integrated intensity [Fig.~\ref{fig2:tdep}(e)] indicates that the (0,1,0) peak appears below $\sim$60~K, and first changes very slowly with temperature, before experiencing an abrupt change with significant hysteresis, typical of a first-order phase transition and mimicking the magnetic susceptibility [Fig.~\subref*{fig0:structure:c}]. The hysteresis lies well below the structural ($T_{S} \approx 60$~K) and magnetic ($T_{N} \approx 58$~K) transitions. In contrast, the integrated intensity of (1,1,0) changes only slightly and with only a very small hysteresis in the 30--50~K range.
	
In addition, we observed $E_{i}$-dependent scattering near (0.5,0.5,0), attributed to double scattering from out-of-plane (0.5,0,0.5) magnetic Bragg peaks, providing an opportunity to probe the magnetic order parameter in the $(H,K,0)$ plane simultaneously with (0,1,0) and (1,1,0), which would have been otherwise inaccessible. As in the NPD data [Fig.~\subref*{fig0:structure:d}--\subref{fig0:structure:f}], the integral intensity exhibits no hysteresis, in contrast with peak intensity.
	
\begin{figure}[tpb]
\subfloat{
\includegraphics[width=0.95\linewidth]{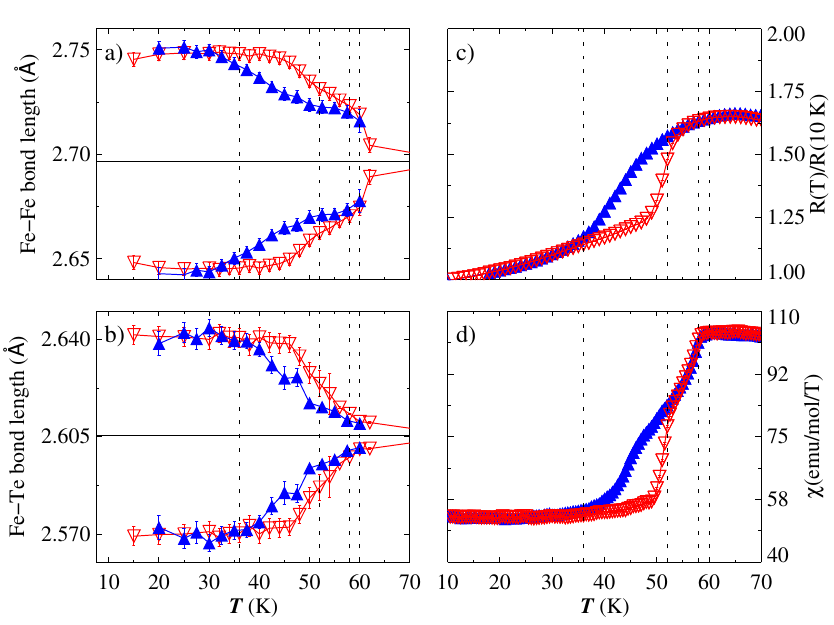}
\label{fig3:bonds:a}}
\subfloat{\label{fig3:bonds:b}}\subfloat{\label{fig3:bonds:c}}\subfloat{\label{fig3:bonds:d}}
\vspace{-0.1in}
\caption{Fe--Fe (a) and Fe--Te (b) bond length in Fe$_{1.09(1)}$Te obtained by fitting the (1,0,0) and (1,1,0) intensities in the data similar to that in Fig. \ref{fig2:tdep} on cooling (closed) and warming (open symbols) to model Eq. (\ref{Eq-1}). Resistivity (c) and magnetic susceptibility (d) measured on cooling (closed) and warming (open symbols) in the same Fe$_{1+y}$Te, $y = 0.09(1)$ sample. Formation of Fe--Fe zigzag chains manifests itself by the concomitant hysteretic decrease in both quantities.}
\label{fig3:bonds}
\vspace{-0.1in}
\end{figure}

Here we fit the temperature dependence data to a simple structure factor model where we have introduced small displacements of Fe (along $a$) and Te (along $b$) from their high symmetry positions in the $a$--$b$ plane,
\begin{eqnarray}
\label{Eq-1}
|F_{(100)}|^{2}&=&4 [ b_{Fe} \sin(2\pi\delta_{x}^{Fe})]^{2}, \\
|F_{(010)}|^{2}&=&4 [ b_{Te} \sin(2\pi\delta_{y}^{Te})]^{2}, \nonumber
\end{eqnarray}
where $b_{Fe}$ and $b_{Te}$ are the neutron scattering lengths and $\delta_{x}^{Fe}$ and $\delta_{y}^{Te}$ are the atomic displacements in r.l.u. The observed broad structure of the (1,0,0)/(0,1,0) and (1,1,0) peaks results from $a$--$b$ twinning in our crystal, combined with the presence of displacements both along $a$ and $b$ directions, suggesting a low-temperature space group with lower symmetry than $P2_{1}/m$, $e.g.~Pc$, \cite{supplement}. Although other lower symmetry groups may also be considered, the slight difference between them is in the stacking of the $a$--$b$ planes, an unessential feature undetectable in powder diffraction \cite{Li2009,Bao2009,Martinelli2010}. The essential common aspect of these space groups, unequal Fe--Fe bonds forming zigzag patterns, Fig. \ref{fig0:structure:a}, is still described by Eq. \ref{Eq-1}.

In Fig.~\ref{fig3:bonds} we present the resulting temperature dependence of the bond lengths between Fe--Fe nearest-neighbors [Fig.~\subref*{fig3:bonds:a}] and Fe--Te nearest-neighbors [Fig.~\subref*{fig3:bonds:b}]. The data indicate that the Fe--Fe bonds shorten (and correspondingly lengthen) from $\sim$2.70~\AA, the bond lengths when atoms are at their high-symmetry positions, forming the 1D zigzag chains illustrated in Fig.~\subref*{fig0:structure:b}, a modulation pattern corresponding to a bond-order wave \cite{Yuan2002}, while the corresponding Fe--Te bonds lengthen and shorten respectively. The Fe--Fe bond modulation is consistent with previous NPD reports in the monoclinic phase \cite{Zaliznyak_PRB2012,Bao2009,Martinelli2010,Rodriguez2011}; the key feature here is the temperature dependence where the dramatic growth of bond disparity shows the same hysteresis as resistivity [Fig.~\subref*{fig3:bonds:c}].
Comparing resistivity to magnetic susceptibility we note that no significant changes in conductivity occur near the structural ($T_{S}$) or magnetic ordering ($T_{N}$) transitions, indicating that the observed hysteretic transition is likely purely electronic in origin, driven by changes in hybridization with temperature, which induce a BOW consistent with ferro-orbital order.

This conclusion is further supported by the large magnitude of bond modulation, $\approx 0.1$~\AA, consistently obtained from both NPD and single crystal data. While bond disparity can also result from lattice- and magneto-striction in association with structural or magnetic transition, as has been anticipated for the case of bicollinear order \cite{Paul_PRB2011}, experimental examples indicate small effects. A bond length disparity of $\approx 0.01$~\AA\ occurs at the structural transition in Fe$_{1.01}$Se \cite{McQueen_PRL2009}. In BaFe$_2$Se$_3$, Fe-Se ladders have a bond modulation of 0.22~\AA\ associated with plaquette orbital order, which  increases by just $\approx 0.03$~\AA\ due to magnetic ordering \cite{Nambu_PRB2012,Caron_PRB2012}. Hence, the bond disparity observed here indicates a substantial modulation of the orbital character in the bonds.

The effects of this orbital order on the magnetic state are apparent. At this intermediate $y$, helimagnetic order competes with the bicollinear commensurate magnetism, and, in fact, incommensurate order appears first with the decreasing temperature [Figs.~\subref*{fig0:structure:d},\subref{fig0:structure:e}]. However, the changing hybridization leads to the formation of the ferromagnetic zigzag chains illustrated in Fig.~\subref*{fig0:structure:b}, which stabilize the bicollinear order, causing a shift of scattering away from an incommensurate position $(\delta,0,0.5)$ towards $Q(0.5,0,0.5) \approx 0.96$~\AA$^{-1}$. The total ordered magnetic moment is unchanged upon cooling/warming, but since the hybridization is hysteretic, and controls the shift in weight between the two competing orders, we observe a hysteresis in the peak intensity, but not the integrated intensity of the magnetic order parameter.

These results present a clear picture, allowing the reconciliation of a number of recent experiments and predictions. ARPES measurements on Fe$_{1.02}$Te revealed a sharp feature near the Fermi energy $E_{F}$ below $T_{N}$, suggesting the appearance of coherent charge carriers in the AFM phase \cite{Liu_Shen2013}. Simultaneously, neutron scattering shows a decrease of local moments from spin-3/2 at high temperature to spin-1 at low temperature \cite{Zaliznyak_PRL2011}. In the light of our observations, these phenomena reflect a change in the character of charge carriers near $E_{F}$ and indicate charge delocalization within the emerging ferromagnetic 1D zigzag Fe chains.

An electronic decoherence-coherence crossover near $E_{F}$ in FeTe upon cooling in this temperature range has been recently predicted by DMFT calculations \cite{Yin_PRB2012}. Additionally, orbital ordering forming 1D Fe chains has been predicted to be a key ingredient for bicollinear order in FeTe \cite{Turner2009}, supported by a quantitative Wannier-function analysis \cite{WeiKu2010}. Furthermore, these results bear some similarity with a recent theoretical study which suggests proximity in FeSC end members to a nematic state arising from a breaking of $C_{4}$ symmetry, driven by hybridization \cite{StanevLittlewood2013}.

In summary, we have presented a detailed study of Fe$_{1+y}$Te ($y = 0.09(1)$), revealing a distinctive first-order bond order wave transition, separate from the magnetic and structural symmetry breakings, and consistent with ferro-orbital ordering. The hysteretic temperature dependence of specific Bragg peaks observed in neutron scattering, correlated with similar hysteresis in magnetic susceptibility and resistivity, suggests the presence of an electronically driven transition. By applying a simple structural model to neutron scattering data, we have mapped the displacements of the Fe and Te atoms from their high symmetry positions, revealing significant splitting of the in-plane Fe--Fe bond lengths. This suggests temperature-dependent hybridization leading to orbital order forming 1D zigzag Fe chains. %

Our results call for a fundamental revision of the paradigm where only structural and magnetic transitions are considered important players in the phase diagram of unconventional
superconductors, and firmly establish that BOW and orbital ordering associated with temperature-dependent electronic coherence must be taken into account in such correlated-electron systems.

\begin{acknowledgments}
Work at BNL was supported by Office of Basic Energy Sciences, US DOE, under Contract No. DE-AC02-98CH10886. Research conducted at ORNL's High Flux Isotope Reactor and Spallation Neutron Source was sponsored by the Scientific User Facilities Division, Office of Basic Energy Sciences, US Department of Energy. We acknowledge the support of NIST, US Department of Commerce, in providing the neutron research facilities used in this work.
\end{acknowledgments}

\emph{Note added}: Two studies that have just appeared \cite{Koz_PRB2013,Rodriguez_PRB2013} confirm the first-order character of the lower-T transition.


%

\end{document}